\documentclass[aps,prb,twocolumn,floatfix,showpacs,superscriptaddress]{revtex4}
\usepackage{graphicx}
\usepackage{bbm}
\usepackage{amsmath,amssymb}

\newcommand{\nc}{\newcommand}
\newcommand{\be}{\begin{equation}}
\newcommand{\beq}{\begin{equation}}
\newcommand{\ee}{\end{equation}}
\newcommand{\bea}{\begin{eqnarray}}
\newcommand{\eea}{\end{eqnarray}}
\newcommand{\ba}{\begin{array}}
\newcommand{\ea}{\end{array}}
\newcommand{\nn} {\nonumber}
\nc{\inv}[1]{\frac{1}{#1}}
\nc{\pa}{\partial}
\nc{\na}{\nabla}
\nc{\goto}{\rightarrow}
\nc{\lgoto}{\longrightarrow}
\nc{\Goto}{\Rightarrow}
\nc{\lGoto}{\Longrightarrow}
\renewcommand{\vr} {{\bf r}}
\newcommand{\vk} {{\bf k}}
\newcommand{\vy} {{\bf y}}
\newcommand{\vR} {{\bf R}}
\nc{\al}{\alpha}
\nc{\bet}{\beta}
\nc{\g}{\gamma}
\nc{\Del}{\Delta}
\nc{\e}{\epsilon}
\nc{\eps}{\epsilon}
\nc{\lam}{\lambda}
\nc{\Lam}{\Lambda}
\nc{\om}{\omega}
\nc{\Om}{\Omega}
\nc{\ve}{\varepsilon}
\nc{\vp}{\varphi}

\begin{document}
\title{On the Kirzhnits gradient expansion in two dimensions}
\author{A. Putaja}
\affiliation{Nanoscience Center, Department of Physics, University of
  Jyv\"askyl\"a, FI-40014 Jyv\"askyl\"a, Finland}
\author{E. R{\"a}s{\"a}nen}
\email[Electronic address:\;]{erasanen@jyu.fi}
\affiliation{Nanoscience Center, Department of Physics, University of
  Jyv\"askyl\"a, FI-40014 Jyv\"askyl\"a, Finland}
\affiliation{Physics Department, Harvard University, 02138 Cambridge MA, USA}
\author{ R. van Leeuwen}
\affiliation{Nanoscience Center, Department of Physics, University of
  Jyv\"askyl\"a, FI-40014 Jyv\"askyl\"a, Finland}
\author{J. G. Vilhena}
\affiliation{Universit\'e de Lyon, F-69000 Lyon, France and
LPMCN, CNRS, UMR 5586, Universit\'e Lyon 1, F-69622 Villeurbanne, France}
\author{M. A. L. Marques}
\affiliation{Universit\'e de Lyon, F-69000 Lyon, France and
LPMCN, CNRS, UMR 5586, Universit\'e Lyon 1, F-69622 Villeurbanne, France}

\date{\today}

\begin{abstract}
We derive the semiclassical Kirzhnits expansion of the
$D$-dimensional one-particle density matrix up to the second order 
in $\hbar$. We focus on the two-dimensional (2D) case and show
that all the gradient corrections both to the 
2D one-particle density and to the kinetic energy 
density vanish. However, the 2D Kirzhnits expansion satisfies the 
consistency criterion of Gross and Proetto
[J. Chem. Theory Comput. {\bf 5}, 844 (2009)] for the functional
derivatives of the density and the noninteracting 
kinetic energy with respect to the Kohn-Sham potential.
Finally we show that the gradient correction to the exchange 
energy diverges in agreement with the previous linear-response 
study.
\end{abstract}


\maketitle 

\section{Introduction}

Gradient expansions provide a natural path to correct 
the local-density approximation (LDA) for slowly-varying
densities as already suggested by
Hohenberg and Kohn in their seminal paper on
density-functional theory~\cite{hohenberg} (DFT).
The second-order gradient expansion for the
kinetic energy was shown to be very useful, 
but on the other hand, systematic
gradient expansions for the exchange and 
particularly for the 
correlation energies faced problems
that were later corrected -- at least
from the practical viewpoint -- by generalized-gradient
approximations~\cite{perdewkurth} (GGAs).
Nevertheless, gradient expansions pose
still open questions, especially
in reduced dimensions such as the two-dimensional
electron gas~\cite{vignale} (2DEG). The interest in the 2DEG
arises from a multitude of applications in, e.g.,
quantum Hall and semiconductor physics.

Semiclassical gradient expansions can be 
regarded as alternatives to the standard approaches based
on Taylor expansions and linear-response formalism.
Although semiclassical methods do not give access to
the correlation energy, they can be used to derive
simple density functionals for the Kohn-Sham (KS) kinetic
energy $T_s$ and the exchange energy density 
$\epsilon_x$ (Ref.~\onlinecite{grossbook}).
Here we focus on the semiclassical Kirzhnits commutator
formalism.~\cite{kirts} It has been previously used to derive the
lowest-order (second order in $\hbar$) gradient correction 
terms to the one-particle density matrix $\gamma(\vr,\vr')$ and 
to $\epsilon_x$ in three 
dimensions (3D),~\cite{grosspaper} as well as for $T_s$
in $D$ dimensions.~\cite{salasnich} Higher-order
corrections in 3D have been considered in Ref.~\onlinecite{googlebook}.
Salasnich~\cite{salasnich} found that in 2D the 
gradient corrections to the kinetic energy vanish, 
which was in agreement with earlier results based on
the response-function approach.~\cite{holas}
This 2D gradient correction was also recently studied 
for systems at finite temperature.~\cite{arxiv}

In this paper we use the Kirzhnits method to derive the 
lowest-order gradient corrections to the one-particle 
density matrix in $D$ dimensions. Then we focus on the 2D
case and show that all the corrections
to the one-particle density $n(\vr)$ vanish, and, in agreement
with Ref.~\onlinecite{salasnich} they vanish also
for $T_s$. Due to the resulting simple expressions
for $n(\vr)$ and $T_s$
the consistency criterion of Gross and 
Proetto~\cite{consistency} that couples 
the functional derivatives of $T_s$ and $n(\vr)$ 
is trivially satisfied.
Finally, we show that
the gradient corrections to $\epsilon_x$ {\em diverge} 
in the 2D Kirtzhnits expansion, 
which is in agreement with the linear-response 
results of Gumbs and Geldart.~\cite{gumbs}

\section{Kirzhnits expansion in $D$ dimensions}

The exchange energy $E_x$ and the KS kinetic energy $T_s$
can be expressed as~\cite{grossbook,parrbook}
\begin{align}
\label{exchange}
E_{x}&=-\inv{4}\int d^{D}r\,d^{D}r'\frac{\left|\gamma(\vr,\vr')\right|^{2}}{\left|\vr-\vr'\right|}\\
T_{s}&=-\frac{\hslash^{2}}{2m}\int\,d^D r\left\{\na^{2}_{\vr}\gamma(\vr,\vr')\right\}_{\vr'=\vr},
\label{kinetic}
\end{align}
where the one-particle density matrix
can be written in terms of the Fermi energy $\e_{F}$ as
 \begin{align}\label{exact}
 \gamma(\vr,\vr')&=\sum_{j:\e_{j}\leq\e_{F}}\vp_{j}^{*}(\vr)\vp_{j}(\vr')\nn\\
&=\sum_{j}\Theta(\e_{F}-\e_{j})\vp_{j}^{*}(\vr)\vp_{j}(\vr')\nn\\
&=\big<\vr|\Theta(\e_{F}-\hat{t}-\hat{v}_{S})|\vr'\big>.
\end{align}
Here $\varphi_i$ are the solutions of the single-particle 
KS equation, $\hat{t}$ is the kinetic energy operator,
$\hat{v}_{s}$ is the KS potential, and $\varTheta$ is 
the Heaviside step function. 
Now we define the local Fermi energy 
$\hat{E}_{F}\equiv\e_{F}-v_s(\vr)$ 
and use the plane-wave decomposition as
\begin{align}\label{planewave}
 \gamma(\vr,\vr')&=\sum_{\alpha=\pm}\int d^{D}k\big<\vr|\Theta(\hat{E}_{F}-\hat{t})|\vk\alpha\big> \big<\vk\alpha|\vr'\big>,
\end{align}
where $|\vk\alpha\big>$ (with $\alpha$ as the spin index) 
are eigenfunctions of the momentum operator $\hat{p}$.

We introduce the abbreviated
notations: $\Theta(\hat{E}_{F}-\hat{t})=f(\hat{a}+\hat{b})$, $f=\varTheta$, 
$\hat{a}=-\hat{t}=-\hat{p}^{2}/2$, and
$\hat{b}=\hat{E}_{F}=\hat{k}_{F}^{2}/2$. Now 
we can use the inverse Laplace transform, the Fourier-Mellin integral,
to show that the operator $\Theta(\hat{E}_{F}-\hat{t})$ 
acts on eigenfunctions $|\vk\big>$ as
\begin{align}\label{flaplace}
f(\hat{a}+\hat{b})|a\big>&=\mathcal{L}^{-1}\{F(\beta)\}|a\big>\nn\\
&=\inv{2\pi i}\int_{c-i\infty}^{c+i\infty}d\beta F(\beta)e^{\beta(\hat{a}+\hat{b})} |a\big>,
\end{align}
where $c={\rm Re}(\beta)>0$ is arbitrary, but 
chosen such that the contour path of the integration 
is in the region of convergence of $F(\beta)$.
The commutation problem of operators $\hat{a}$ and $\hat{b}$ 
can be avoided by introducing a new operator 
$\hat{K}(\beta)$ (Refs.~\onlinecite{grossbook} and \onlinecite{googlebook})
such that 
\bea
e^{\beta(\hat{a}+\hat{b})}= e^{\beta\hat{b}}\hat{K}(\beta)e^{\beta\hat{a}}.
\label{operatorK}
\eea
Thus, we obtain for Eq.~(\ref{flaplace}) an expression
\begin{align}
f(\hat{a}+\hat{b})|a\big>&=\inv{2\pi i}\int_{c-i\infty}^{c+i\infty}d\beta F(\beta)e^{\beta(a+\hat{b})}\hat{K}(\beta)|a\big>,
\end{align}
where the operator $\hat{a}$ has now been 
replaced with eigenvalue $a$ in the exponential 
function, so that it commutes with the operator $\hat{b}$.

The expression for the operator $\hat{K}(\beta)$ 
is obtained by expanding it in a power series with 
respect to $\beta$,
\begin{align}\label{kexpand}
 \hat{K}(\beta)=\sum_{n=0}^{\infty}\beta^{n}\hat{O}_{n}.
\end{align}
Differentiating both sides of Eq.~(\ref{operatorK}) 
with respect to $\beta$, 
and expanding all exponential functions in a Taylor series,
leads to the recurrence relation~\cite{grossbook,googlebook}
\begin{align}
\hat{O}_{0}&=1, \qquad \hat{O}_{1}=0,\\
\hat{O}_{n+1}&=\inv{n+1}\left([\hat{a},\hat{O}_{n}]+\sum_{j=1}^{n}\hat{C}_{j}\hat{O}_{n-j}\right)\\
\hat{C}_{j}&=\frac{(-1)^{j}}{j!}\underbrace{[\hat{b},[\hat{b},[...[\hat{b}}_{ j\textrm{ times}},\hat{a}]...].
\end{align}
Inserting Eq. (\ref{kexpand}) in Eq.~(\ref{flaplace}) yields
\begin{align}
\label{kirzhnits}
 f(\hat{a}+\hat{b})|a\big> &=\sum_{n=0}^{\infty}\biggl[\inv{2\pi i}\int_{c-i\infty}^{c+i\infty}d\beta F(\beta)\beta^{n}e^{\beta(a+\hat{b})}\biggl]\hat{O}_{n}|a\big>\nn\\
&=\sum_{n=0}^{\infty}f^{(n)}(a+\hat{b})\hat{O}_{n}|a\big>,
\end{align}
where $f^{(n)}$ is the n:th derivative of the function $f$. 
The Kirzhnits expansion in
Eq.~(\ref{kirzhnits}) leads to the following 
expression of the density matrix,
\begin{align}\label{seriesdensity}
&\gamma(\vr,\vr')=\underbrace{\sum_{\alpha=\pm}\int d^{D}k\,\Theta\left(E_{F}
-\frac{k^{2}}{2}\right)\big<\vr|\vk\alpha\big>\big<\vk\alpha|\vr'\big>}_{\g^{(0)}}\nn\\
&+\sum_{n=2}^{\infty}\underbrace{\sum_{\alpha=\pm}\int d^{D}k\,\delta^{(n-1)}\left[E_{F}
-\frac{k^{2}}{2} \right]\big<\vr|\hat{O}_{n}|\vk\alpha\big>\big<\vk\alpha|\vr'\big>}_{B_{n}},
\end{align}
where $\delta^{(n)}$ is the n:th derivative of the delta function. 
The first-order term of Eq.~(\ref{seriesdensity})
$\g^{(0)}$ corresponds to the zeroth order solution of the 
one-particle density matrix and it can be written as
\begin{align}\label{zerothg}
 \g^{(0)}&=\sum_{\alpha=\pm}\int d^{D}k\,\Theta\left(E_{F}-\frac{\hslash^{2}k^{2}}{2m}\right)\big<\vr|\vk\alpha\big>\big<\vk\alpha|\vr'\big>\nn\\
&=\frac{2\delta_{\sigma,\sigma'}}{(2\pi)^{D}}\int_{0}^{k_{F}}dk\,k^{D-1}\underbrace{\int d\Omega e^{iky\cos\theta}}_{\equiv I(k)}\nn\\
&=\frac{2\delta_{\sigma,\sigma'}}{(2\pi)^{D}}\int_{0}^{k_{F}}dk\,k^{D-1}I(k),
\end{align}
where $d\Omega$ is the $(D-1)$-dimensional angular volume element
and $\theta$ is the angle between vectors $\vy$ and $\vk$.
This term generates the exact exchange energy for the 
homogeneous electron gas which can be used as the LDA 
in an inhomogeneous system.
Note that in Eq.~(\ref{zerothg}) we have defined the
relative and center-of-mass coordinates as $\vy=\vr-\vr'$ 
and ${\bf R}=(\vr+\vr')/2$, respectively.

Higher-order terms of the Kirzhnits expansion $\g^{(n)}$ 
can be determined by calculating higher derivatives of the delta function
and multiple commutators of $\hat{E}_{F}$ and $\hat{t}$ that lead to 
multiple derivatives of $k_{F}$. 
The second-order ($\na^{2}$) inhomogeneity correction 
consists of three terms, $\g^{(2)}(\vr,\vr')=B_{2}+B_{3}+B_{4}$, 
where
\begin{align}
   B_{2}&=\frac{\delta_{\sigma,\sigma'}}{(2\pi)^{D}}\biggl[\na^{2}_{\vR}k_{F}^{2}f(z)+2(\na_{\vR} k_{F}^{2})\cdot\na_{\vy}f(z)\biggr]\nn\\
&=\frac{\delta_{\sigma,\sigma'}}{(2\pi)^{D}}\biggl[\na^{2}_{\vR}k_{F}^{2}f(z)+2 k_{F}\frac{\pa f}{\pa z}(\na_{\vR} k_{F}^{2})\cdot\frac{\vy}{y}\biggr]\\
B_{3}&=\frac{2\delta_{\sigma,\sigma'}}{3(2\pi)^{D}}\biggl\{2\na^{2}_{\vR}k_{F}^{2}\frac{k_{F}^{2}}{z}\frac{\pa g}{\pa z}+
2k_{F}^{2}\na_{\vR}\left[(\na_{\vR} k_{F}^{2})\cdot\frac{\vy}{y}\right]\cdot\frac{\vy}{y}\nn\\
&\times\left(\frac{\pa^{2}g}{\pa z^{2}}
-\inv{z}\frac{\pa g}{\pa z}\right)+(\na_{\vR}k_{F}^{2})^{2}g(z)\biggr\}\\ 
B_{4}&=\frac{\delta_{\sigma,\sigma'}}{(2\pi)^{D}}\biggl\{(\na_{\vR}k_{F}^{2})^{2}\frac{k_{F}^{2}}{z}\frac{\pa h}{\pa z}
+k_{F}^{2}\left[(\na_{\vR}k_{F}^{2})\cdot\frac{\vy}{y}\right]^{2}\nn\\
&\times\left(\frac{\pa^{2}h}{\pa z^{2}}-\inv{z}\frac{\pa h}{\pa z}\right)\biggr\}.
\end{align}
Here we use a definition $z=z(\vR,y)=k_{F}(\vR)|\vy|$ and expressions
\begin{align}
 f(z)&=\int\,d^{D}\vk\,\delta'\left[E_{F}-\inv{2}k^{2}\right]e^{i\vk\cdot\vy}\nn\\
&=\frac{k_{F}^{d-4}}{4}\left[(d-2)I(z)+z I'(z)\right];\\
g(z)&=\int\,d^{D}\vk\,\delta''\left[E_{F}-\inv{2}k^{2}\right]e^{i\vk\cdot\vy}\nn\\
&=\frac{k_{F}^{d-6}}{8}\biggl\{\left[d^{2}-6d+8\right]I(z)+(2d-5)z I'(z)\nn\\
&+z^{2}I''(z)\biggr\};\\
h(z)&=\int\,d^{D}\vk\,\delta'''\left[E_{F}-\inv{2}k^{2}\right]e^{i\vk\cdot\vy}\nn\\
&=\frac{k_{F}^{d-8}}{16}\biggl\{\left[d^{3}-12d^{2}+44d-48\right]I(z)\nn\\
&+3(d^{2}-7d+11)z I'(z)+3(d-3)z^{2}I''(z)\nn\\
&+z^{3}I^{(3)}(z)\biggr\}.
\end{align}

Combining our results leads to the semiclassical expansion 
of the density matrix of the form
\begin{align}
\g(\vr,\vr')&=\g^{(0)}(\vr,\vr')+\g^{(2)}(\vr,\vr')\nn\\
&=\delta_{\sigma,\sigma'}\biggl\{A+B(\na_{\vR}k_{F}^{2})\cdot\frac{\vy}{y}
+C\,\na^{2}_{\vR}k_{F}^{2}\nn\\
&+D\,\na_{\vR}\left[(\na_{\vR}k_{F}^{2})\cdot\frac{\vy}{y}\right]\cdot\frac{\vy}{y}
+E\,(\na_{\vR}k_{F}^{2})^{2}\nn\\
&+F\left[(\na_{\vR}k_{F}^{2})\cdot\frac{\vy}{y}\right]^{2}
\biggr\},
\end{align}
where $A$, $B$, $C$, $D$, $E$, and $F$ are given by
\begin{align}
A&=\frac{2}{(2\pi)^{D}}\int_{0}^{k_{F}}dk\,k^{d-1}I(k)\nn\\
B&=\frac{2 k_{F}}{(2\pi)^{D}}\frac{\pa f}{\pa z}\nn\\
C&=\frac{1}{(2\pi)^{D}}\left(f(z)+\frac{4}{3}\frac{k_{F}^{2}}{z}\frac{\pa g}{\pa z}\right)\nn\\
D&=\frac{4 k_{F}^{2}}{3(2\pi)^{D}}\left(\frac{\pa^{2}g}{\pa z^{2}}-\inv{z}\frac{\pa g}{\pa z}\right)\nn\\
E&=\frac{1}{(2\pi)^{D}}\left(\frac{2}{3}g(z)+\frac{k_{F}^{2}}{z}\frac{\pa h}{\pa z}\right)\nn\\
F&=\frac{k_{F}^{2}}{(2\pi)^{D}}\left(\frac{\pa^{2}h}{\pa z^{2}}-\inv{z}\frac{\pa h}{\pa z}\right).\nn
\end{align}
Using these equations it is straightforward to proceed with
the calculation of the one-particle density matrix and the 
exchange energy density in 2D.

\section{Two-dimensional case}

\subsection{One-particle density matrix and
the kinetic energy}

In the two-dimensional case we obtain the 
following expression for the remaining integral
in the one-particle density matrix in 
Eq.~(\ref{zerothg}),
\begin{align}
I(k)=\int d\Omega e^{iky\cos\theta}=2\pi J_{0}(z).
\end{align}
This leads to
\begin{align}\label{2dmatrix}
 \g(\vr,\vr')&=\g^{(0)}+\g^{(2)}(\vr,\vr')\nn\\
&=\frac{\delta_{\sigma,\sigma'}}{\pi}\biggl\{k_{F}^{2}\frac{J_{1}(z)}{z}
-\inv{4}z J_{0}(z)\inv{k_{F}}(\na_{\vR} k_{F}^{2})\cdot\frac{\vy}{y}\nn\\
&-\inv{24}z J_{1}(z)\frac{\na^{2}_{\vR} k_{F}^{2}}{k_{F}^{2}} \nn\\
&+\inv{12}z^{2}J_{0}(z)\inv{k_{F}^{2}}\na_{\vR}\left((\na_{\vR}k_{F}^{2})\cdot\frac{\vy}{y}\right)
\cdot\frac{\vy}{y}\nn\\
&+\inv{96}z^{2}J_{2}(z)\frac{(\na_{\vR}k_{F}^{2})^{2}}{k_{F}^{4}}\nn\\
&-\inv{32}z^{3}J_{1}(z)\inv{k_{F}^{4}}\left((\na_{\vR}k_{F}^{2})\cdot\frac{\vy}{y}\right)^{2}
\biggr\},
\end{align}
where $J_{n}(z)$ is the Bessel function of the 
first kind in order $n$. This expression is 
the central result of this paper and in the
following it is used for further analysis.

We first notice that the one-particle density has
a simple form
\begin{align}\label{density}
 n(\vr)=\inv{2\pi}k_{F}^{2}(\vr),
\end{align}
i.e., all the gradient corrections vanish. 
This may seem as an unexpected result in view
of the known gradient expression in 3D.~\cite{grosspaper}
Secondly, we calculate the KS kinetic energy 
by inserting Eqs.~(\ref{2dmatrix}) and (\ref{density}) into 
Eq.~(\ref{kinetic}) and find
\begin{align}\label{kineticenergy}
 T_{s}&=\int d\vr \,t_{s}(\vr)\nn\\
&=-\frac{\hslash^{2}}{2m}\int\,d\vr\left\{\na^{2}_{\vr}\g_{S}(\vr,\vr')\right\}_{\vr'=\vr}\nn\\
&=-\frac{\hslash^{2}}{2m}\int\,d\vR\left\{\left(\frac{1}{2}\na_{\vR}+\na_{\vy}\right)^{2}\g(\vR,y)\right\}_{y=0},
\end{align}
where $t_{s}$ is the noninteracting kinetic energy density. 
After lengthy but straightforward calculations
we find
\beq\label{kineticdensity}
t_{s}(\vr)=\frac{\hslash^{2}}{2m}\pi\,n^{2}(\vr),
\ee
which is equal to the Thomas-Fermi expression.
Again, the gradient correction (von Weizs\"acker term)
is zero. This 2D result is in agreement with 
previous Kirzhnits expansion for the $D$-dimensional
kinetic energy~\cite{salasnich} as well as with
results obtained using alternative 
methods.~\cite{holas,shao,koivisto}

In their recent work Proetto and Gross~\cite{consistency} have
derived a rigorous condition to test the consistency of 
approximations made for the density and the KS kinetic 
energy. The condition is given by
\begin{align}\label{test}
 \frac{\delta T_{s}[v_{s}]}{\delta v_{s}(\vr)}=-\int d\vr' v_{s}(\vr')\frac{\delta n[v_{s}](\vr')}{\delta v_{s}(\vr)},
\end{align}
where $v_s$ is the KS potential. 
We note that the condition follows from the 
Euler equation minimizing the KS energy, i.e.,
\be
\frac{\delta T_s}{\delta n(\vr')}=-v_s(\vr')+\mu,
\ee 
where $\mu$ is the chemical potential. Multiplying
both sides with $\delta n(\vr') / \delta v_s(\vr)$
and integrating over $\vr'$ directly yields Eq.~(\ref{test}).
The condition means that 
$\delta T_{s}[n]/\delta n(\vr')=\epsilon_{F}-v_{s}$ must
be also valid. Using Eqs.~(\ref{density}) and (\ref{kineticdensity}), 
and $k_{F}=\sqrt{2m(\epsilon_{F}-v_{s})/\hslash^{2}}$ 
we find
\begin{align}
 \frac{\delta T^{\rm TF}[n]}{\delta n(\vr')}&=\frac{\hslash^{2}}{m}\pi\,n=\frac{\hslash^{2}}{2m}k_{F}^{2}=\epsilon_{F}-v_{s}.
\end{align}
Thus, Eq.~(\ref{test}) is fulfilled for the 2D (and also 3D)
results of the semiclassical Kirzhnits expansion.

\subsection{Exchange energy}

Knowledge of the gradient corrections to the one-particle 
density matrix in Eq.~(\ref{2dmatrix}) immediately 
motivates to search for an expression for the exchange
energy defined in Eq.~(\ref{exchange}). Using
Green's first theorem we obtain the second-order 
expansion of the exchange energy density in $\hslash$ 
in terms of the gradients of $k_{F}$:
\begin{align}
e_{x}(\vr)&=-\inv{4}\int d^{2}\vr'\frac{|\gamma(\vr,\vr')|^{2}}{|\vr-\vr'|}\nn\\
&=-\frac{2}{3\pi^{2}}k_{F}^{3}\nn\\
&-\inv{192\pi}\frac{(\na k_{F}^{2})^{2}}{k_{F}^{3}}\int\,dz\,D(z),
\end{align}
where
\begin{align}
D(z)&=z^{2}J_{0}(z)^{2}+(z^{2}-4)J_{1}(z)^{2}-2zJ_{1}(z)J_{2}(z)\nn\\
&+4zJ_{0}(z)J_{1}(z)+z^{2}J_{0}(z)J_{2}(z).
\end{align}
Expanding Bessel functions in Taylor series and using a 
regularization of divergent Coulomb integrals leads 
to 
\begin{align}
 \lim_{\alpha\rightarrow 0}&\int_{0}^{\infty}dz e^{-\alpha z}D(z)\nn\\
&=\Bigl/_{0}^{\infty}\biggl\{\frac{2}{3}z^{3}-\frac{1}{8}z^{5}\nn\\
&+\frac{5}{672}z^{9}-\frac{7}{675840}z^{11}+
\frac{121}{191692800}z^{13}-\ldots\biggr\}\nn\\
&=\infty.
\end{align}
In other words, the exchange energy density is
clearly divergent in the 2D Kirzhnits expansion.
Our result agrees with the finding of Gumbs
and Geldart~\cite{gumbs} who used
perturbation theory and linear-response formalism
to derive the second-order gradient terms
for both the kinetic and exchange energies
in $D$ dimensions. They arrived at the same result
also by using the Wigner-Kirkwood 
expansion.~\cite{gumbs2}
Hence, as confirmed in this work from the
semiclassical point of view, the divergence of 
the systematic gradient expansion for the exchange 
energy seems to be an inevitable mathematical fact.
However, to the best of our knowledge, the underlying 
{\em physical} reason that makes the 2D situation 
specially divergent in contrast with the 1D and 3D 
cases remains unknown. We hope that the present
analysis encourages further examinations from that
viewpoint.

The divergence of the exchange energy in 2D 
can be considered unfortunate in view of 
functional developments in 2D, although 
first GGAs in 2D have already been 
obtained,~\cite{gga} and several other
2D functionals have been derived, for example, 
in the framework of meta-GGAs.~\cite{metagga}
A natural next step, as already discussed
in Ref.~\onlinecite{gumbs}, would be 
considering expansions in quasi-2DEG by
introducing a finite width of the system.
This would resemble also the experimental
situation in low-dimensional nanostructures
such as in semiconductor quantum dots.

\section{Summary}

In summary, we have derived the second-order
gradient corrections to the one-particle
density matrix in the semiclassical Kirzhnits 
expansion in $D$ dimensions. In two dimensions
the corrections vanish in the diagonal of the 
density matrix, i.e., in the one-particle
density. Similar vanishing occurs in the
noninteracting kinetic energy in agreement
with Ref.~\onlinecite{salasnich}, and leads to
the fulfillment of the consistency criterion
of Ref.~\onlinecite{consistency}.
Finally, we have shown that the exchange
energy of the two-dimensional Kirzhnits expansion
diverges in agreement with the 
linear-response theory. We hope that
the present work motivates further 
attempts in the systematic derivation
of gradient corrections in the
quasi-two-dimensional electron gas.

\begin{acknowledgments}
This work was supported by the 
Academy of Finland (A.P. and E.R.) and the
Wihuri Foundation (E.R.). 
G.V. acknowledges support from the FCT 
(Grant No. SFRH/BD/38340/2007) and
M.A.L.M. from the 
French ANR (ANR-08-CEXC8-008-01).
We are grateful to
E.K.U. Gross and C.R. Proetto for valuable 
discussions.
\end{acknowledgments}

\end{document}